%%%%%%%%%%%%%%%%%%%%%%%%%%%%%%%%%%%%%%%%%%%%%%%%%%%%%%%%%
%%%%%%%%%%%%%%%%%%%%%%%%%%%%%%%%%%%%%%%%%%%%%%%%%%%%%%%%%
%%                                                     %%
%%                (Plain) TeX file for                 %%
%%                                                     %%
%%      An  Infra-Red Finite Electron Propagator       %%
%%                                                     %%
%%                         by                          %%
%%                                                     %%
%%        E. Bagan, M. Lavelle and D. McMullan         %%
%%                                                     %%
%%                                                     %%
%%             to appear in PRD, 8/July/97             %%
%%                                                     %%
%%%%%%%%%%%%%%%%%%%%%%%%%%%%%%%%%%%%%%%%%%%%%%%%%%%%%%%%%
%%%%%%%%%%%%%%%%%%%%%%%%%%%%%%%%%%%%%%%%%%%%%%%%%%%%%%%%%

  %First some fonts

\font\bigbold=cmbx12
\font\eightrm=cmr8
\font\sixrm=cmr6
\font\fiverm=cmr5
\font\eightbf=cmbx8
\font\sixbf=cmbx6
\font\fivebf=cmbx5
\font\eighti=cmmi8  \skewchar\eighti='177
\font\sixi=cmmi6    \skewchar\sixi='177
\font\fivei=cmmi5
\font\eightsy=cmsy8 \skewchar\eightsy='60
\font\sixsy=cmsy6   \skewchar\sixsy='60
\font\fivesy=cmsy5
\font\eightit=cmti8
\font\eightsl=cmsl8
\font\eighttt=cmtt8
\font\tenfrak=eufm10
\font\sevenfrak=eufm7
\font\fivefrak=eufm5
\font\tenbb=msbm10
\font\sevenbb=msbm7
\font\fivebb=msbm5
\font\tensmc=cmcsc10
\font\tencmmib=cmmib10  \skewchar\tencmmib='177
\font\sevencmmib=cmmib10 at 7pt \skewchar\sevencmmib='177
%\font\sevencmmib=cmmib7 \skewchar\sevencmmib='177
\font\fivecmmib=cmmib10 at 5pt \skewchar\fivecmmib='177
%\font\fivecmmib=cmmib5 \skewchar\fivecmmib='177
%Some Families

\newfam\bbfam
\textfont\bbfam=\tenbb
\scriptfont\bbfam=\sevenbb
\scriptscriptfont\bbfam=\fivebb

\newfam\frakfam
\textfont\frakfam=\tenfrak
\scriptfont\frakfam=\sevenfrak
\scriptscriptfont\frakfam=\fivefrak

\newfam\cmmibfam
\textfont\cmmibfam=\tencmmib
\scriptfont\cmmibfam=\sevencmmib
\scriptscriptfont\cmmibfam=\fivecmmib
\def\bold#1{\fam\cmmibfam\relax#1}

%Definition of 8 point

\def\eightpoint{%
\textfont0=\eightrm   \scriptfont0=\sixrm
\scriptscriptfont0=\fiverm  \def\rm{\fam0\eightrm}%
\textfont1=\eighti   \scriptfont1=\sixi
\scriptscriptfont1=\fivei  \def\oldstyle{\fam1\eighti}%
\textfont2=\eightsy   \scriptfont2=\sixsy
\scriptscriptfont2=\fivesy
\textfont\itfam=\eightit  \def\it{\fam\itfam\eightit}%
\textfont\slfam=\eightsl  \def\sl{\fam\slfam\eightsl}%
\textfont\ttfam=\eighttt  \def\tt{\fam\ttfam\eighttt}%
\textfont\bffam=\eightbf   \scriptfont\bffam=\sixbf
\scriptscriptfont\bffam=\fivebf  \def\bf{\fam\bffam\eightbf}%
\abovedisplayskip=9pt plus 2pt minus 6pt
\belowdisplayskip=\abovedisplayskip
\abovedisplayshortskip=0pt plus 2pt
\belowdisplayshortskip=5pt plus2pt minus 3pt
\smallskipamount=2pt plus 1pt minus 1pt
\medskipamount=4pt plus 2pt minus 2pt
\bigskipamount=9pt plus4pt minus 4pt
\setbox\strutbox=\hbox{\vrule height 7pt depth 2pt width 0pt}%
\normalbaselineskip=9pt \normalbaselines
\rm}

%More general stuff

\def\pagewidth#1{\hsize= #1}
\def\pageheight#1{\vsize= #1}
\def\hcorrection#1{\advance\hoffset by #1}
\def\vcorrection#1{\advance\voffset by #1}

\newcount\notenumber  \notenumber=1              %Numbering does
\newif\iftitlepage   \titlepagetrue              %not start on title
\newtoks\titlepagefoot     \titlepagefoot={\hfil}%page
\newtoks\otherpagesfoot    \otherpagesfoot={\hfil\tenrm\folio\hfil}
\footline={\iftitlepage\the\titlepagefoot\global\titlepagefalse
           \else\the\otherpagesfoot\fi}

\def\abstract#1{{\parindent=30pt\narrower\noindent\eightpoint\openup
2pt #1\par}}
\def\smc{\tensmc}

%A nicer footnote

\def\note#1{\unskip\footnote{$^{\the\notenumber}$}
{\eightpoint\openup 1pt
#1}\global\advance\notenumber by 1}

\def\frac#1#2{{#1\over#2}}
\def\dfrac#1#2{{\displaystyle{#1\over#2}}}

\def\({\left(}
\def\){\right)}
\def\<{\langle}
\def\>{\rangle}
   %Partial derivatives
\def\2pd#1#2#3{\frac{\partial^2#1}{\partial#2\partial#3}}

\def\sqr#1#2{{\vcenter{\vbox{\hrule height.#2pt
        \hbox{\vrule width.#2pt height#1pt \kern#1pt
           \vrule width.#2pt}
        \hrule height.#2pt}}}}

\def\ni{\noindent}

\def\slash{\!\!\!/\,}

%%% Macro to generate the equation #'s automatically.
%%% To use start each new section (eg 3) with the commands
%%% \secno=3 \meqno=1 :this will start the equations with (3.1)
%%% Then in place of \eqno(3.1) type \eqn\descriptivename . To refer
%%% back to the equation simply type (\descritivename)
%%% For the appendix set \secno=0, \appno=1\meqno=1 etc
%%% If there are no sections, then set \secno=0

\global\newcount\secno \global\secno=0
\global\newcount\meqno \global\meqno=1
\global\newcount\appno \global\appno=0
\newwrite\eqmac
\def\romappno{\ifcase\appno\or A\or B\or C\or D\or E\or F\or G\or H
\or I\or J\or K\or L\or M\or N\or O\or P\or Q\or R\or S\or T\or U\or
V\or W\or X\or Y\or Z\fi}
\def\eqn#1{
        \ifnum\secno>0
            \eqno(\the\secno.\the\meqno)\xdef#1{\the\secno.\the\meqno}
          \else\ifnum\appno>0
            \eqno({\rm\romappno}.\the\meqno)\xdef#1{{\rm\romappno}.
               \the\meqno}
          \else
            \eqno(\the\meqno)\xdef#1{\the\meqno}
          \fi
        \fi
\global\advance\meqno by1 }

\def\eqnn#1{
        \ifnum\secno>0
            (\the\secno.\the\meqno)\xdef#1{\the\secno.\the\meqno}
          \else\ifnum\appno>0
            \eqno({\rm\romappno}.\the\meqno)\xdef#1{{\rm\romappno}.
                \the\meqno}
          \else
            (\the\meqno)\xdef#1{\the\meqno}
          \fi
        \fi
\global\advance\meqno by1 }
%%% Macro to assist in the references
%%% At the beginning of the paper list the references in the order
%%% that they appear by the command \refn
%%% So if the first reference is to be
%%%  D. McMullan and I. Tsutsui Nucl. Phys. B121 (1994) 12
%%% then type \refn\us{D. McMullan and I. Tsutsui\np{121}{94}{12}}
%%% In the text this is simply referred to by [\us].
%%% At the end of the text type \listrefs

\global\newcount\refno
\global\refno=1 \newwrite\reffile
\newwrite\refmac
\newlinechar=`\^^J
\def\ref#1#2{\the\refno\nref#1{#2}}
\def\nref#1#2{\xdef#1{\the\refno}
\ifnum\refno=1\immediate\openout\reffile=refs.tmp\fi
\immediate\write\reffile{
     \noexpand\item{[\noexpand#1]\ }#2\noexpand\nobreak.}
     \immediate\write\refmac{\def\noexpand#1{\the\refno}}
   \global\advance\refno by1}
\def\semi{;\hfil\noexpand\break ^^J}
\def\nl{\hfil\noexpand\break ^^J}
\def\refn#1#2{\nref#1{#2}}

\def\up#1{$^{[#1]}$}

\def\cmp#1#2#3{{\it Commun. Math. Phys.} {\bf {#1}} (19{#2}) #3}
\def\jmp#1#2#3{{\it J. Math. Phys.} {\bf {#1}} (19{#2}) #3}
\def\ijmp#1#2#3{{\it Int. J. Mod. Phys.} {\bf A{#1}} (19{#2}) #3}

\def\pl#1#2#3{{\it Phys. Lett.} {\bf {#1}B} (19{#2}) #3}
\def\np#1#2#3{{\it Nucl. Phys.} {\bf B{#1}} (19{#2}) #3}

\def\pr#1#2#3{{\it Phys. Rev.} {\bf {#1}} (19{#2}) #3}

\def\prD#1#2#3{{\it Phys. Rev.} {\bf D{#1}} (19{#2}) #3}
\def\prl#1#2#3{{\it Phys. Rev. Lett.} {\bf #1} (19{#2}) #3}

\def\ann#1#2#3{{\it Ann. Phys.} {\bf {#1}} (19{#2}) #3}
\def\prp#1#2#3{{\it Phys. Rep.} {\bf {#1}C} (19{#2}) #3}
\def\tmp#1#2#3{{\it Theor. Math. Phys.} {\bf {#1}} (19{#2}) #3}
\def\zpC#1#2#3{{\it Z. Phys.} {\bf C{#1}} (19{#2}) #3}

\def\fortschr#1#2#3{{\it Fortschr. d. Phys.} {\bf {#1}} (19{#2}) #3}

\def\cjp#1#2#3{{\it Can. J. Phys.} {\bf #1} (19{#2}) #3}

\def\empty#1#2#3{{\bf{#1}} (19{#2}) #3}

%% Some Macros specific to this note%%

\def\a{\alpha}
\def\b{\beta}

\def\eps{\epsilon}

\def\psic{\psi_{\rm c}}

\def\d{\delta}

\def\p{\pi^0}

\def\ket#1{\vert#1\rangle}
\def\pa{\partial}

\def\vdenom#1#2{{\left[{(#1_1-#2_1)^2}\gamma^{2}+ (#1_2-#2_2)^2+
(#1_3-#2_3)^2\right]}}

\def\vv{{\bold v}}
\def\al{\alpha}
\def\be{\beta}
\def\cG{{\cal G}}
\def\cI{{\cal I}}
\def\cov{(p-k)^2-m^2}

\def\de{\delta}
\def\De{\Delta}
\def\Dx{ {dx\over \sqrt{1-x}\sqrt{1-\vv{}^2 x}}}

\def\ep{\epsilon}
\def\ga{\gamma}
\def\integ{\int       {   d^{2\om} k\over  (2\pi)^{2\om}    }       }
\def\noncov{k^2-(k\cdot\eta)^2+(k\cdot v)^2}
\def\om{\omega}
\def\p{\partial}
\def\pn{p\cdot\eta}
\def\pv{p\cdot v}
\def\slash{\!\!\!/\,}

%%% Some parameters for this note %%%

\pageheight{24cm}
\pagewidth{15.5cm}
%\hcorrection{-2.5mm}
\magnification \magstep1
\voffset=8truemm
\baselineskip=16pt
\parskip=5pt plus 1pt minus 1pt

%%Figures macro%%

\input epsf.tex
\def\fig#1#2{$$\epsfxsize=#2\epsfbox{#1}$$}
%%% The equations

\secno=0

%%% The references
{\eightpoint
\refn\HATFIELD{B.\
Hatfield, {\sl Quantum Field Theory of Point Particles
and Strings}, (Addison-Wesley, Redwood City 1992)}
\refn\TARRACH{R.\ Tarrach, \np{183}{81}{384}}
\refn\TOM{J.C.\ Breckinridge, M.\ Lavelle and T.G.\ Steele,
\zpC{65}{95}{155}}
\refn\COHONE{P.\ Kulish and L.\ Faddeev, \tmp{4}{70}{745}}
\refn\JAUCH{J.M.\ Jauch and F.\ Rohrlich, {\it The Theory
of Photons and Electrons}, Second Expanded
Edition, (Springer-Verlag, New York 1980)}
\refn\STROCCHI{F.\ Strocchi and A.S.\ Wightman, \jmp{15}{74}{2198}}
\refn\ZWANZIG{D.\ Maison and D.\ Zwanziger, \np{91}{75}{425}}
\refn\MONSTER{M. Lavelle and D. McMullan, \prp{279}{97}{1}}
\refn\HAAG{R.\ Haag, {\sl Local Quantum Physics}, (Springer-Verlag,
Berlin, Heidelberg, 1993)}
\refn\BUCHHOLZ{D.\ Buchholz, \pl{174}{86}{331}}
\refn\FACTSNFIC{D.\ Buchholz, \np{469}{96}{333}}
\refn\SPINNER{R.L.\ Jaffe, \pl{365}{96}{359}}
\refn\CORNW{J.M.\ Cornwall, {\sl private communication}}
\refn\WEIN{S. Weinberg, \prl{31}{73}{494}}
\refn\GROS{D.J. Gross and F. Wilczek, \prD{8}{73}{3633}}
\refn\GELL{H. Fritzsch. M. Gell-Mann and H. Leutwyler,
\pl{47}{73}{269}}
\refn\QCDONE{M.\ Belloni, L.\ Chen and K.\ Haller, \prD{55}{97}{2347}}
\refn\MID{L.\ Lusanna, \ijmp{10}{95}{3675}}
\refn\QCDLAST{P.\ Haagensen and K.\ Johnson, {\sl On the Wavefunctional 
for Two Heavy Colour Sources in Yang-Mills Theory}, hep-th/9702204}
\refn\DIRACINCANADA{P.A.M.\ Dirac, \cjp{33}{55}{650}}
\refn\MANDEL{S.\ Mandelstam, \ann{19}{62}{1}}
\refn\STEINMANN{O.\ Steinmann, \ann{157}{84}{232}}
\refn\STEINMANNAGAIN{O.\ Steinmann, \np{350}{91}{355};
\empty{361}{91}{173}}
\refn\EMILIOE{E.\ d'Emilio and M.\ Mintchev, \fortschr{32}{84}{473},
503}
\refn\BUCHSTRING{D.\ Buchholz, \cmp{85}{82}{49}}
\refn\SHAB{See, e.g., http://www.ifae.es/\~{}roy/qed.html and 
S.V.\ Shabanov, {\sl The Proper Field of Charges and
Gauge-Invariant Variables in Electrodynamics}, Dubna preprint,
E2-92-136 (unpublished)}
\refn\DIRAC{P.A.M. Dirac, {\sl Principles of Quantum Mechanics},
(OUP, Oxford, 1958)}
\refn\SYMA{K. Symanzik, {\sl Lectures on Lagrangian Quantum Field
Theory}, DESY preprint T-71/1 (unpublished)}
\refn\LETTER{E.\ Bagan, M.\ Lavelle and D.\ McMullan, \pl{370}{96}{128}}
\refn\LORBOOK{W.K.H.\ Panofsky and M.\ Phillips, {\sl Classical 
Electricity and Magnetism}, (Addison-Wesley, Reading Ma.\ 1962)}
\refn\JACKIW{R.\ Jackiw, \prl{41}{78}{1635}}
\refn\ADKINS{G.S.\ Adkins, \prD{27}{83}{1814}}
\refn\JACKSOL{R.\ Jackiw and L.\ Soloviev, \pr{173}{68}{1485}}
\refn\COLOUR{M.\ Lavelle and D.\ McMullan, \pl{371}{96}{83}} 
\refn\PANP{M. Lavelle and D. McMullan, \pl{329}{94}{68}}
\refn\SSB{M.\ Lavelle and D.\ McMullan, \pl{347}{95}{89}}
%\refn\AXIAL{See, e.g., {\sl Physical and Non-Standard Gauges}, ed.'s
%P.\ Gaigg et al, Springer Lecture Notes in Physics {\bf 361}
%(Springer-Verlag, Heidelberg 1990)}
}
%
%the beginning
%
%\leftline{{\bigbold DRAFT VERSION}}
%\rightline{??BNL??}
\rightline {UAB-FT-384}
\rightline {PLY-MS-96-01}
\vskip 22pt
\centerline{\bigbold AN INFRA-RED FINITE ELECTRON PROPAGATOR}
\vskip 20pt
\centerline{\smc Emili Bagan{\hbox
{$^1$}}\footnote{{*}}{{\eightpoint\rm Current and permanent
address:
IFAE, Universitat Aut\`onoma de Barcelona.}},
Martin Lavelle{\hbox {$^2$}}
and  David McMullan{\hbox {$^3$}}}
\vskip 15pt
{\baselineskip 12pt
\centerline{\null$^1$Physics Department}
\centerline{Bldg.\ 510A}
\centerline{Brookhaven National Laboratory}
\centerline{Upton, NY 11973}
\centerline{USA}
\centerline{email: iftebag@cc.uab.es}
\vskip 13pt
\centerline{\null$^2$Grup de F\'\i sica Te\`orica and IFAE}
\centerline{Edificio Cn}
\centerline{Universitat Aut\`onoma de Barcelona}
\centerline{E-08193 Bellaterra (Barcelona)}
\centerline{Spain}
\centerline{email: lavelle@ifae.es}
\vskip 13pt
\centerline{\null$^{3}$School of Mathematics and Statistics}
\centerline{University of Plymouth}
\centerline{Drake Circus, Plymouth, Devon PL4 8AA}
\centerline{U.K.}
\centerline{email: d.mcmullan@plymouth.ac.uk}}
\vskip 1truemm
\vskip 12pt
{\baselineskip=13pt\parindent=0.58in\narrower\ni{\bf Abstract}\hskip
2truemm
We investigate the properties of a dressed electron which reduces, in a
particular class of gauges, to the usual fermion.
A one loop calculation of the propagator is presented.
We show explicitly that an infra-red finite,
multiplicative, mass shell renormalisation is possible for this dressed
electron, or, equivalently, for the usual fermion in the
abovementioned gauges. The results are
in complete accord with previous conjectures.
\par}
\bigskip\bigskip
\centerline{\it To appear in Physical Review D}

\vfill\eject
\noindent
{\bf 1) Introduction}
\smallskip
\ni A fundamental question in gauge theories is:
what is the correct description of an asymptotic field?
In an abelian theory this problem takes on 
its most pristine form, and the
obstacle to adopting the naive in-out identification of
asymptotically free fields is clearly identified with the infra-red
divergences associated with the  masslessness
of the gauge fields.
As such, we will restrict 
ourselves in this paper to Quantum Electrodynamics (QED), i.e., a
non-confining, abelian
gauge theory where the gauge symmetry is unbroken.
After presenting and heuristically motivating
an Ansatz for a charged particle (henceforth an \lq
electron\rq) in QED, this description will be put
to a highly non-trivial test: we will calculate the
one-loop propagator and show that it is infra-red finite in a suitable
(and previously predicted)
mass-shell renormalisation scheme. Another
interpretation of this result is that we have found a new class of
gauges, parameterised by a vector ${\bold v}$, where the usual fermion
propagator is infra-red finite.

Mass shell renormalisation of the electron propagator is hindered in
most gauges by the 
appearance of infra-red divergences (see, e.g., p.\
410 of Ref.\ \HATFIELD) although the position of the pole is itself
gauge independent\up{\TARRACH,\TOM}.
It is well known that these infra-red problems are a
consequence of the difficulties in defining the physical
asymptotic fields correctly. In the confining theory of Quantum
Chromodynamics (QCD) this is self-evidently a highly non-trivial
problem, but even in our paradigm theory, perturbative QED, no
satisfactory answer has yet been given to this question. It is
understood that the masslessness of the photon means that the
electromagnetic interaction falls off too slowly for us to just ignore
it and replace the physical electron by a bare fermion. The
coherent state technique\up{\COHONE}, where one adds soft
photons, has been
developed to deal with these divergences. For a summary
of the usual approaches we refer to Supplement 4 of Ref.\ \JAUCH.
Despite this
understanding of the root of the infra-red problem, it does not seem
that a full description of charged states in gauge theories exists. The
coherent state approach has not, for example,
been carried through for the strong
interaction. However, even for QED previous work on
dressing electrons 
seems somewhat {\it ad-hoc} and prescriptive in nature.
In what follows we will stress 
the systematic and predictive nature of the approach we advocate.

There are certain general properties to be found in
 any description of an
electron: it must be non-local\up{\STROCCHI-\MONSTER}
and it must be non-covariant\up{\MONSTER-\BUCHHOLZ}. Both these
things follow from the gauge symmetry of QED. Non-locality can be
simply shown to follow from demanding that Gauss' law holds on a
physical, gauge invariant state, a more rigorous
proof is contained in Ref.\ \STROCCHI. The
non-covariance of such a description is a result of the difficulties in
reconciling Lorentz and gauge symmetries in the charged sector (see
Sect.\ 8 of Ref.\ \MONSTER). At the naivest level these requirements
amount to the need to dress a charge with an electromagnetic \lq
cloud\rq, whose exact form depends upon the position and velocity of
the charge. The neglect of such a dressing when one uses a bare fermion
as an asymptotic field is equivalent to switching off the coupling
which is clearly unphysical and this in fact
underlies the infra-red problem.

Although these divergences may, however, be, essentially,
ignored in calculations
of scattering processes in QED it is clear that a better understanding
of their origins and of how to describe physical charged states
is highly desirable. An understanding of bound states cannot come from 
switching off the coupling, even asymptotically, and insight into how 
to dress the constituent charges of, e.g., positronium would, we feel, 
be of great practical value. 
Furthermore, in QCD, which is worse affected by
such infra-red problems, the asymptotic region is really the short
distance regime\up{\FACTSNFIC} and so an
understanding of the dressings associated
with colour charges will yield valuable information about the gluons
and sea-quarks in hadrons --- our present lack of understanding of this
structure being revealed most glaringly in the so-called proton spin
crisis (see, e.g., Ref.\ \SPINNER). We remark that dressings underlie 
Cornwall's pinch technique\up{\CORNW} and also recall here the
long-suggested connection between the severe infra-red divergences of
QCD and the confinement phenomenon\up{\WEIN-\GELL}. Further attempts 
to construct gauge invariant descriptions of quarks and gluons may be 
found in Ref.'s \QCDONE-\QCDLAST.

How should we now dress our electron? We expect to surround the charge
with a cloud and, since the dressed particle should correspond to a
physical state, we expect our expression to be gauge invariant. Many
years ago Dirac presented such a formula\up{\DIRACINCANADA}:
$$
\psi_{\rm f}(x)=\exp\(-ie\int d^4z f^\mu(x-z)A_\mu (z)
\)\psi(x)
\,,
\eqn\fdesc
$$
where $f_\mu$ is a field-independent function obeying
$$
\pa_\mu f^\mu(w)=\d^{(4)}(w)
\,,
\eqn\no
$$
and we note that the sign of $e$ used in this paper is the opposite 
to that of Bjorken and Drell. 
It may be straightforwardly seen that this is gauge invariant. It is
also visibly non-local and, depending upon the choice of $f^\mu$, can
be non-covariant. Several
authors have employed this formula (see, e.g., Ref.'s
\MANDEL-\BUCHSTRING) to study
the construction of physical states. The stability of such dressings
around static charges in QED was considered in Ref.\ \SHAB.

We now note that there is a gauge in which the argument of the
exponential in (\fdesc) vanishes: $f^\mu(x) A_\mu(x)=0$, we will
call such gauges \lq dressing gauges\rq.
This connection between a specific type of gauge fixing and the dressing
for a charged state is quite general and explained in more detail in
Ref.\ \MONSTER. This simple observation, 
however, has an important consequence
for us: we expect that if one dresses the charge correctly no
infra-red problem will arise. We now see that working in the dressing
gauge should also permit an infra-red finite mass shell
renormalisation if the dressing is a physical one.
In the light of the known general structures associated with any
construction of an electron, we still need to make the form of 
$f_\mu$, and hence the particular dressing gauge, precise.

Our first restriction is to limit the form of the non-locality of the
cloud. In Ref.\ \MONSTER\ it was argued that one must avoid non-locality
in time otherwise there would be no natural prescription for the
identification of asymptotic fields for the far distant past and future.
One can, in principal, have a dressing that is local in time outside
some bounded interval of time. However, for the class of dressings we
are interested in here, we restrict the dressing to  a
particular time-slice, i.e., we assume that, $f_0=0$.
This specification notwithstanding, we
still have a great deal of freedom in our choice of the
three $f_i$-components.

The next step is to recall that Dirac (see Ref.\ \DIRACINCANADA\
and Sect.\ 80 of Ref.\ \DIRAC)
suggested using the following form for the $f_i$:
$$
\psi_{\rm c}(x)=\exp\(-ie\frac{\pa_iA_i}{\nabla^2}(x)\)\psi(x)
\,,
\eqn\staticdesc
$$
where the action of $\nabla^{-2}$ is understood as
$$
\frac1{\nabla^2}g(x_0,{\bold x})=-\frac1{4\pi}\int d^3z
\frac{g(x_0,{\bold z})}{\vert {\bold x-z}\vert}
\,.
\eqn\no
$$
It is clear that this is a special case of (\fdesc) and is hence gauge
invariant. The dressing gauge here is the familiar Coulomb gauge. The
appealing feature of this choice of dressing is that
the commutators of the electric and magnetic fields with (\staticdesc)
yield just the electric and magnetic fields we expect of a static
charge. Using the canonical equal time commutator,
$[E_i({\bold x}),A_j({\bold y})]=i\d_{ij}\d({\bold x-y})$, one finds,
for example, that taking an
eigenstate $\ket{\eps}$ of the electric field
operator, with eigenvalue $\eps_i$, and adding a dressed fermion
(\staticdesc) to the system then
$$
E_i({\bold x})\psic({\bold y})\ket\eps=
\left( \eps_i({\bold x}) -
\frac{e}{4\pi}\frac{{\bold x}_i-{\bold y}_i}{\vert
{\bold x-y}\vert^3} \right) \psic({\bold y})\ket\eps
\,,
\eqn\statE
$$
This means that it is natural to interpret this dressed, gauge
invariant fermion as describing a static charge.

It might now be argued that this last argument, based as it is on the
free field canonical commutation relations and hence completely
ignoring renormalisation, may not hold in the full theory.
For this reason two of us recently\up{\MONSTER}
considered the one-loop propagator of
the dressed charge (\staticdesc) in a general covariant gauge and in
Coulomb gauge\note{It may appear that the above description is local in
Coulomb gauge, recall, however, that in that gauge we must use Dirac
brackets and the bracket between the fermion and the electric field is
non-local. See Ref.'s \SYMA\ and \MONSTER\ for details.}.
The results demonstrated
that a multiplicative, infra-red finite, mass shell
renormalisation of the propagator was possible. It was, however, only
possible at the static mass shell point, $p=(m,0,0,0)$ --- which 
is of course
in complete accord with the above interpretation of this dressing.

Although this result is highly attractive and sheds new light on the
infra-red finiteness of the Coulomb gauge, it covers
in some sense only \lq one point\rq\ in a space of dressings. In Ref.\
\MONSTER\ a gauge invariant description of a dressed charge moving with
some constant velocity, which reduces in the static limit to
(\staticdesc), was presented (see Sect.\ 2 below for the specific
form of this dressing). It was there conjectured that the
propagator of this dressed electron would be infra-red finite if the
correct (moving) renormalisation 
point on the mass shell was used. In a recent letter\up{\LETTER}
we demonstrated that, in the small velocity limit, a multiplicative
renormalisation of this Ansatz was possible. No new infra-red
divergences arose, but it was clear that this could be the case when
terms of order ${\bold v}^2$ were retained in the dressing. In this
paper we will consider the dressed propagator for an arbitrary velocity
and verify the conjectures of Ref.\ \MONSTER. The usual 
electron propagator
will, in other words, be shown to be infra-red finite in a class of
gauges depending upon a free parameter (the three-vector, ${\bold v}$).

After this introduction, the rest of this paper is structured as
follows. In Sect.\ 2 we discuss the exact form of the dressing we
use and the equivalent (dressing) gauge. We also describe the
renormalisation of the fermion propagator in different gauges in QED.
Sect.~3, the heart of the paper, is devoted to the explicit
regularisation and renormalisation of the propagator. Here we obtain
the promised result that an infra-red finite mass shell renormalisation
is possible. In Sect.\ 4 a discussion of our results is presented.
An appendix devoted to the integrals we have required
concludes this work.
\bigskip
\goodbreak
\ni {\bf 2) The Dressing, the Gauge and the Self-Energy}
\smallskip
\ni The dressed electron
which we will work with in this paper has the
form\up{\MONSTER,\LETTER}
$$
\eqalign{
\psi_{{\bold v}}=& \exp\Bigg(\Bigg.
\frac{ie}{4\pi}\gamma\cr
&\times\int d^3z\frac{\gamma^{-2}
\pa_1A_1(x^0,{\bold z})+\pa_2A_2(x^0,{\bold
z})+\pa_3A_3(x^0,{\bold z})-v^1 E_1(x^0,{\bold
z})}{\vdenom{x}{z}^{\frac12}}\Bigg.\Bigg) \psi(x)\,,
}
\eqn\thedressing
$$
where $\gamma=1/\sqrt{1-{\bold v}^2}$ and ${\bold v}=(v^1,0,0)$. 
We propose it for the following reasons: it is gauge invariant and
its commutators with the electric and magnetic fields are such that
$$
{\bold E}(x)=-\frac{e}{4\pi}\gamma\frac{{\bold x}-{\bold
y}}{\vdenom{x}{y}^{\frac32}}\,,
\eqn\no
$$
and
$$
{\bold B}(x)={\bold v}\times{\bold E}(x)\,,\eqn\no
$$
which one may recognise as the correct electric and magnetic fields for
a charge moving with constant velocity, ${\bold v}$, 
along the $x_1$-axis (see, e.g., Chap.\ 19 of Ref.\ \LORBOOK).
This expression is analogous to (\staticdesc)
and indeed reduces to it for ${\bold v}\to 0$.
In the non-relativistic case this dressed electron reduces to
$$
\psi_{\bold v}(x)=
\exp\left(
-ie\frac{\pa_jA_j+ v_i E_i}{\nabla^2}
\right)\psi(x)
\,.
\eqn\smallv
$$
The renormalisation of the propagator of this field at order $e^2$ and
first order in ${\bold v}$ is to be found in Ref.\ \LETTER. Before 
computing the propagator of $\psi_{\bold v}$, we will now briefly 
discuss its complex relation with that of the static dressed electron, 
$\psic$. 

It is important to first 
note that the form of the dressing appropriate to the
moving electron (\thedressing) does not follow from a naive boost to the
dressing for the static electron (\staticdesc). This is a concrete
manifestation of the fact\up{\MONSTER-\BUCHHOLZ} that Lorentz
transformations cannot be
implemented unitarily on charged fields. As such, it is not possible to
argue that the good infra-red properties found in the static case can be
simply boosted up to the moving dressing. Given the surprising nature of
this fact, it is helpful to show how such a boost must act on such a
charged field and hence make clear why it is not now a unitary mapping.

We recall that as a four vector, the potential $A_\mu(x)$ transforms under
a Lorentz transformation $x\to x'=\Lambda x$
as ${A_\mu}'(x)=UA_\mu(x)U^{-1}$ where $U$ is
the appropriate unitary operator and ${A_\mu}'(x)=\Lambda_\mu^\nu A_\nu(x')$.
Under a boost with velocity, ${\bold v}$, in
the $x^1$ direction we find that the dressing gauge appropriate to the
static charge becomes:
$$
\eqalign{
\pa_iA_i(x)\to\gamma^2&\(\gamma^{-2}\pa_1A_1+\pa_2A_2+
\pa_3A_3-v^1E_1\)(x')\cr
&+{\bold v}^2\gamma^2\(\pa_1A_1-\pa_0A_0-\pa_2A_2-\pa_3A_3\)(x')
}\eqn\no
$$
From (\thedressing) we see that the first term in this expression is 
the dressing gauge for the moving charge and so the second term here 
obstructs the identification of the 
dressing gauge that we need for our non-static charge. 
Since we know that we can construct the dressing directly from the 
gauge, this exemplifies the fact that 
on charged states the 
Lorentz transformations are not implemented by the
unitary mapping, $U$. 
However, as argued in Refs.\ \JACKIW\ and \MONSTER,
a gauge covariant implementation of the Lorentz transformations can be
constructed by combining the above unitary transformation with a field
dependent gauge transformation. Thus, to transform the static dressing
to the boosted one we take $A_\mu(x)\to \tilde{A}_\mu(x)$, where
$$
\tilde{A}_\mu(x)={A_\mu}'(x)+\pa_\mu\Theta(x)\,,\eqn\no
$$
and
$$
\Theta(x)=\frac{{\bold v}^2\gamma^2}{4\pi}\int
d^3z\frac{(\pa_1A_1-\pa_0A_0-\pa_2A_2-\pa_3A_3)(x_0',{\bold z}')}
{\vert {\bold x}-{\bold z}\vert}\,,
\eqn\no
$$
where the point $(x_0',{\bold z}')$ 
in the integrand is the boost applied to $(x_0,{\bold z})$.

Having constructed the dressing gauge, and hence the 
dressing for a moving charge, we now need to 
address the quantum field theoretic aspects of this approach.
Given the obvious
importance of gauge invariance to us we will work in a gauge invariant
regularisation scheme, viz.\ dimensional regularisation. In consequence
we may drop tadpoles, and we will do this consistently below. 
As a result we can re-express the dressed fermion as
$$
\eqalign{
\psi_{{\bold v}}=& \Bigg(\Bigg.
1-\frac{ie}{4\pi}\gamma\cr
&\times\int d^3z\frac{\gamma^{-2}
\pa_1A_1(x^0,{\bold z})+\pa_2A_2(x^0,{\bold
z})+\pa_3A_3(x^0,{\bold z})-v^1 E_1(x^0,{\bold
z})}{\vdenom{x}{z}^{\frac12}}\Bigg.\Bigg) \psi(x)\,,
}
\eqn\moving
$$
since the $e^2$ terms we so neglect will just yield tadpoles in the 
one-loop 
calculation at hand. This last equation can be rewritten as
$$
\psi_{\bold v}(x)=\left\{1-ie{\ga^{-2}\p_1 A_1+\p_2 A_2+\p_3 A_3
+v^1[\p_0 A_1-\p_1 A_0]
\over \ga^{-2}\p_1^2+\p_2^2+\p_3^2}
 + O(e^2)\right\} \psi(x)\,,
\eqn\moved
$$
where we have employed the standard identity
$$
\left({\p^2\over\p\xi_1^2}+
{\p^2\over\p\xi_2^2}+{\p^2\over\p\xi_3^2}\right)
\left(-{1\over4\pi}\right){1\over\sqrt{\xi_1^2+\xi_2^2+\xi_3^2}}=
\de(\xi_1)\de(\xi_2)\de(\xi_3)\,,
\eqn\no
$$
which under the change of variables, $\xi_i\to\gamma x_i$, 
can be rewritten as
$$
\left({1\over\ga^2}{\p^2\over\p x_1^2}+{\p^2\over\p x_2^2}+
        {\p^2\over\p x_3^2}\right)
\left(-{1\over4\pi}\right){\ga\over\sqrt{\ga^2  x_1^2+ x_2^2+ x_3^2}}=
\de( x_1)\de( x_2)\de( x_3)\,,
\eqn\no
$$
from which (\moved) follows.
It proved in practice convenient to further re-express (\moved)
in a more covariant looking fashion as
$$
\psi_{\bold v}(x)=\left\{1+ie{\cG^\mu A_\mu(x) \over \p^2-
          (\eta\cdot\p)^2+(v\cdot\p)^2} + O(e^2)\right\} \psi(x)\,,
\eqn\covmover
$$
where
$$
\cG_\mu=\left[(\eta+v)_\mu(\eta-v)_\nu -g_{\mu\nu}\right]\pa^\nu
\,,
\eqn\no
$$
and we have introduced the vectors,
$\eta^\mu=(1,0,0,0)$ and
$v^\mu=(0,v^1,0,0)\equiv (0,{\bold v})$ from which the relations
$v\cdot\eta=0$, $\eta^2=1$ and $v^2=-{\bold v}^2$ follow immediately.
We stress that $v$ is {\it not} the four-velocity,
$u^\mu=\ga (1,{\bold v})=\ga(\eta+v)^\mu$.

We may calculate the gauge invariant,
one-loop propagator of $\psi_{\bold v}(x)$ in one
of two ways. One may either work in an arbitrary
Lorentz gauge or one may
perform the calculation in the dressing gauge.
For an arbitrary ${v^1}$ the dressing gauge is now
$$
\ga^{-2}\p_1 A_1+\p_2 A_2+\p_3 A_3
+v^1[\p_0 A_1-\p_1 A_0]= \cG^\mu A_\mu=0
\,,
\eqn\dgauge
$$
and the free photon propagator in this gauge has the form
$$
\eqalign{
D^v_{\mu\nu}=& {1\over k^2} \Big\{\Big. -g_{\mu\nu}
   +{(1-\xi)k^2-[k\cdot(\eta- v)]^2 \ga^{-2}\over [\noncov]^2}
k_\mu k_\nu \cr & \qquad\quad
-{k\cdot(\eta- v)\over\noncov
}\left[k_\mu (\eta+v)_\nu+(\eta+v)_\mu k_\nu\right]
\Big.\Big\}
\,,}
\eqn\dgaugeprop
$$
where $\xi$ is a gauge parameter which we set to zero in what follows;
this ensures, $\cG^\mu D_{\mu\nu}=0$. Even then
this is really a class of gauges parameterised by
${\bold v}$, which flows into the Coulomb gauge for ${\bold v}
\to 0$. We are not aware of any previous work with such gauges.
The Feynman rule for the extra vertex from the dressing is
\medskip
\centerline{
\hbox{
\fig{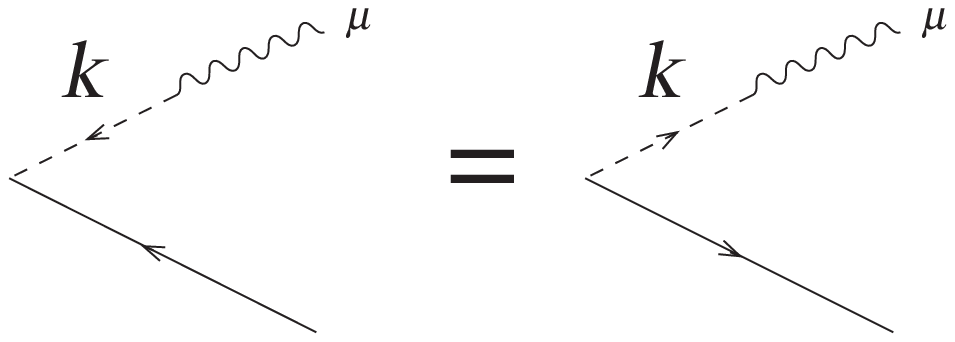}{4.0cm}
%    \vrule height0pt width0.0in
    \vbox{\hbox{$\displaystyle = -e k^\rho\; 
{{ 
(\eta+v)_\mu(\eta-v)_\rho-g_{\mu\rho}  }\over\noncov} \,, $}
 \hrule height0.30cm width0pt}
           }
  }
\medskip\ni With 
these rules we can calculate the dressed propagator in an
arbitrary gauge.

At order $e^2$ we have as well as the usual interaction vertex,
contributions from the expansion in the coupling of the dressing.
These effects mean that, even if we work in a covariant gauge, the
integrand of the sum of all the Feynman diagrams is non-covariant.
We have checked explicitly that,
after discarding tadpoles, the same total integrand
is found in both an arbitrary Lorentz gauge (i.e., it is independent of
the Lorentz gauge parameter) and in the dressing gauge,
(\dgauge). In a general gauge one has to take all of the diagrams of
Fig.\ 1 into account, while in the gauge (\dgauge) only Fig.\ 1a
appears.
\medskip
\midinsert
\fig{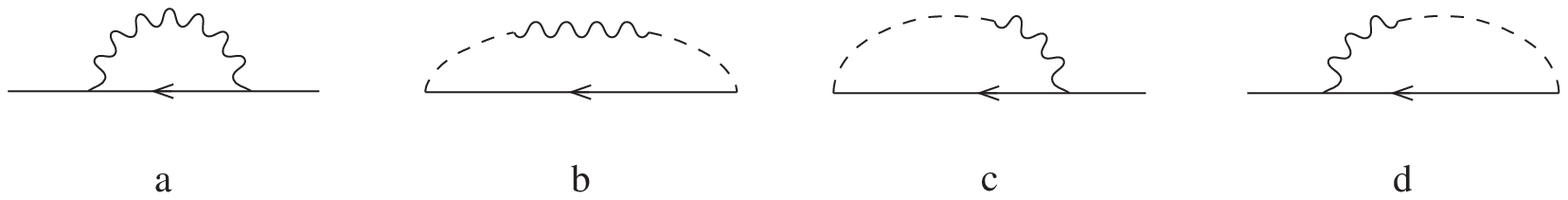}{12cm}
%\centereps{12cm}{5cm}{qf.eps}
\smallskip{\eightpoint\narrower\noindent{\bf Figure 1}
The diagrams which yield the one loop dressed propagator. In the
appropriate dressing gauge only Fig.\ 1.a contributes. In a general
gauge all the diagrams must be evaluated. The dashed lines indicate
the projection of the photon propagator from the
(${\bold v}$-dependent) vertices in the dressing (see the above Feynman 
rules).\par}
\endinsert
The result for the self-energy is (in $D=2\omega$ dimensions)
$$
\eqalign{
-i\Sigma =& e^2 \integ  \Bigg\{\Bigg.
{1\over k^2}  {1\over \cov }
\Big[ 2(\om-1) p\slash - 2\om m -2(\om-1) k\slash \Big]  \cr
+& {1\over k^2}{1\over \cov}   \Big[-2(p\slash -m) \Big] \cr
+&{1\over \cov}{1\over \noncov} \Big[ 2(p\slash-m)+(\eta\slash+v\slash)
             \;  k\cdot (\eta-v)\Big] \cr
+&{1\over \cov}{1\over[ \noncov]^2}(p\slash-m) \Big[
\ga^{-2}(k\cdot\eta-k\cdot v)^2 - k^2 \Big] \cr
+& {1\over k^2}{1\over \cov}{1\over\noncov}\cr
&\quad \times
\Big[-(p^2-m^2)(\eta\slash+v\slash)\; k\cdot(\eta-v)-2k\slash\;
k\cdot(\eta-v)\; p\cdot (\eta+v)\Big] \cr
+& \Bigg. {1\over k^2}{1\over \cov}{1\over[\noncov]^2 }
(p^2-m^2) k\slash \Big[k^2-\ga^{-2}(k\cdot\eta-k\cdot v)^2\Big]
\Bigg\}\,.\!\!\!\!\!\!\!\!\!\!\!\!\!\!\!\!\!\!\!
}
\eqn\selfE
$$
Actually this is the self-energy in the dressing gauge. In covariant
gauges we must include all the diagrams of Fig.\ 1 and so it is more
natural there to consider the whole propagator. For simplicity we will
use the self-energy henceforth.
The detailed
renormalisation of this will be presented in the next section.

\bigskip\goodbreak
\ni {\bf 3) Divergences and Renormalisation}
\smallskip
\ni In this section we will first recall some facts about the
mass-shell renormalisation of the usual fermion propagator and set up
our conventions. We will then give the results of our calculations for
the renormalisation constants.
\bigskip\goodbreak
\ni {\underbar{i) Setting Things Up}}
\smallskip
\ni To renormalise the electron propagator one requires two different
renormalisations: a mass shift ($m\to m-\d m$)
and a fermion wave function
renormalisation. The first of these is known to be gauge independent
and in non-covariant gauges, such as Coulomb gauge, it is independent
of the exact choice of mass shell point (i.e., 
it is the same for all choices
of $p_0$ and ${\bold p}$ which are on-shell)\up{\MONSTER}. 
Based upon our experience with
the renormalisation of the dressed electron (\smallv),
where we only retained terms of first order in ${\bold v}$, we will use
the following multiplicative, matrix renormalisation for the fermion
$$
\psi\to\sqrt{Z_2}\exp\left\{-i\frac{Z'}{Z_2}\sigma_{\mu\nu}
\eta^\mu v^\nu\right\}\psi\,,
\eqn\theZs
$$
which is reminiscent of a naive Lorentz boost upon a fermion.
At lowest order we can recast this as
$$
\psi\to\( \sqrt{Z_2}{\bold I} + \frac{Z'}{\sqrt{Z_2}}
\eta\slash v\slash
\) \psi\,,
\eqn\no
$$
In the small ${\bold v}$ limit such a
multiplicative renormalisation was found to be possible\up{\LETTER}. 
These relations define our three renormalisation constants.
The counterterms in the self-energy can thus be seen to be
(with $Z_2=1+\d Z_2$)
$$
-i\Sigma^{\rm counter}=\d Z_2(p\slash-m)+ 2iZ'(p\cdot\eta v\slash
-p\cdot v\eta\slash)+i\d m\,.
\eqn\cnter
$$
For a multiplicative renormalisation to be possible, the ultraviolet
divergences have to also have this form for arbitrary values of $p^2$,
$p\cdot\eta$, $p\cdot v$ and $v^2$. We find the following such
ultraviolet divergences (see the appendix for a discussion of how to 
perform the integrals) 
$$
\eqalign{
-i\Sigma^{{\rm UV}}=i{\al\over4\pi}{1\over{2-\omega}}\Bigg\{\big.
-3m+&
(p\slash-m)\left[-3-2\chi({\bold v})\right]\cr
&\quad+2(\pv\; \eta\slash-\pn\; v\slash)\left[
{1\over\vv{}^2}+{1+\vv{}^2\over 2 \vv^2}\chi({\bold v})\right]
\big.\Bigg\}\,,
}
\eqn\no
$$
where $\a=(m^2)^{\omega-2}\dfrac{e^2}{4\pi}$  
and we have introduced the definition
$$
\chi({\bold v})=\frac1{\vert 
{\bold v}\vert}{\rm ln}\frac{1-\vert{\bold v}
\vert}  {1+\vert{\bold v}\vert}
\,.
\eqn\chiEq
$$
This displays the need for our matrix multiplication
renormalisation. We note that the UV divergences
are {\it local} in the external
momentum, but {\it non-local} in the velocity $v$.
\bigskip
It is clear that after performing the integrals in (\selfE)
the (renormalised) self-energy including loops and counter terms
will have the general form
$$
-i\Sigma=m\a +p\slash\b +p\cdot\eta\eta\slash \d +mv\slash\eps\,,
\eqn\no
$$
where $\a,\cdots,\eps$
are functions depending upon $p^2, p\cdot \eta, p\cdot v$
and $v^2$. Our choice of renormalisation scheme is to insist that the
on-shell form of the renormalised
propagator is just the tree level one: i.e., there
should be a pole at the physical mass, $m$, and this
should have residue unity. Since the propagator is non-covariant we
must specify for which point on the mass shell 
we will require this. Our
interpretation of this propagator as corresponding to a dressed
electron with velocity given by ${\bold v}$ leads us to choose
the point 
$$
p=m\gamma(1,v^1,0,0)=m\gamma(\eta+v)\,.
\eqn\massShell
$$
The conjecture of Ref.'s \MONSTER\ and \LETTER\ is that the
so-renormalised propagator will be infra-red finite.

To find the mass shift renormalisation constant, $\d m$, we use the
mass-shell condition that there is a pole at $m$. This
implies that the renormalised self-energy must obey  
$$
\tilde \a +\tilde \b +\frac{\( p\cdot\eta\)^2}{m^2}\tilde \d +
\frac{p\cdot v}{m}\tilde \eps=0\,.
\eqn\massShellCond
$$
Here the tildes signify that we put the momentum
$p^2$ on shell in the self-energy (propagator): $p^2=m^2$.
Note that the counterterms, $Z_2$ and $Z'$, from (\cnter) do not enter
in (\massShellCond) since this is on shell and so just $\d m$ will now
be determined.
As stated above, the mass shift is gauge parameter independent
in covariant gauges and it has been seen to be independent of the exact
choice of mass shell point 
in both the Coulomb gauge\up{\MONSTER} and in the
renormalisation of the slowly moving dressed charge\up{\LETTER}. We
therefore expect that (\massShellCond) will hold for {\it any} point on 
the mass shell and this will provide a check on our calculations of the
functions $\a,\cdots,\eps$.

In this notation we may write the Taylor expansion of the
propagator in $(p^2-m^2)$ as
$$
\eqalign{
iS_{\bold v}=i\frac{p\slash+m}{p^2-m^2}-\frac1{p^2-m^2}\Big\{\Big.
(2m^2\tilde \Delta  +\tilde\b)p\slash &
+(2m^2\tilde\Delta +\tilde \a +2\tilde \b)m\cr & -
p\cdot\eta
\eta\slash\tilde \d-m v\slash\tilde\eps\Big.\Big\} +O[(p^2-m^2)^0]
\,,}
\eqn\proppy
$$
where
$$
\tilde \Delta(p\cdot\eta, p\cdot v,v^2) =
\Big(\frac{\pa\a}{\pa p^2} + \frac{\pa\b}{\pa p^2}
+\frac{(p\cdot\eta)^2}{m^2}\frac{\pa\d}{\pa p^2} +\frac{p\cdot v}{m}
\frac{\pa\eps}{\pa p^2}\Big)\Bigm|_{p^2=m^2}
\,.
\eqn\XZX
$$
Note that the infra-red divergences that arise are contained in the
function, $\Delta$.
Clearly we will now require the second term in (\proppy)
to vanish at our renormalisation
point. Requiring that the coefficients of $m$, $\eta\slash$
and $v\slash$ all so vanish at our physically motivated mass shell 
condition (\massShell) gives us three independent equations, which
we choose to write as
$$
\eqalign{
2m^2\bar\Delta+\bar\be-\bar\de&=0\,,\cr
\ga (2m^2\bar\De+\bar\be)-\bar\ep&=0\,,\cr
2m^2\bar\De+\bar\al+2\bar\be&=0\,,
}
\eqn\thecrucialthree
$$
where the bars denote that the functions are now evaluated at $p=\gamma
m (\eta+v)$.

Since we confidently expect the mass shift to be fixed by
Eq.\ \massShellCond\ above, we
seem to have three equations (i.e., Eq.\ \thecrucialthree) and two
unknowns ($\d Z_2$ and $Z'$) and one might worry that perhaps no
solution exists.
However, we can rapidly see that no such problem exists for our choice
of mass shell point. If 
we now explicitly separate out the contributions of
the $\d Z_2$ and $Z'$
counterterms to the self-energy from the rest (and give what is
left, i.e., those coming from the loop integrations and the mass
shift counterterm, a subscript $L$) then we find that
(\thecrucialthree) can be rewritten as
$$
\eqalign{
i\de Z_2-2\vv{}^2\;iZ'&=\bar\de_L-\bar\be_L-2m^2\bar\De\,,\cr
i\de Z_2-
\phantom{\vv{}^2} 2\;iZ'&=\ga^{-1} \bar\ep_L-\bar\be_L-
2m^2\bar\De\,,\cr
i\de Z_2\;\,\phantom{-2\vv{}^2\;iZ'}&=-\bar\al_L-2\bar\be_L-2m^2
\bar\De\,.
}
\eqn\no
$$
We point out that $\Delta=\Delta_L$, i.e., no counterterms appear in
$\Delta$. This set of equations has a solution if
$$
\gamma^2 \bar\d_L+\bar\a_L+\bar\b_L-\gamma {\bold v}^2\bar\eps_L=0
\,,
\eqn\no
$$
and we recognise that this is nothing else but (\massShellCond)
at the physical renormalisation point (\massShell). 
We therefore have the following two equations which
determine our counterterms
$$
\eqalign{
Z'=&{1\over2i}\left[\ga^2 \bar\de_L-\ga\bar\ep_L\right],\cr
Z_2=&-{1\over i}\left[\bar\al_L+2\bar\be_L+2m^2\bar\De \right]\,.
}
\eqn\no
$$

\bigskip\goodbreak
\ni {\underbar{ii) The Renormalisation Constants}}
\smallskip
\ni The calculation of the self-energy and the counterterms is a
laborious task\note{Both MATHEMATICA and REDUCE were used.}.
A discussion of the necessary integrations may be found in the
appendix. Here we will quote the relevant results.
For (\massShellCond) we obtained
$$
\eqalign{
\tilde\al+& \tilde\be+{(p\cdot\eta)^2\over m^2}\tilde\de+
{p\cdot v\over m}\tilde\epsilon=
-i{\al\over4\pi}\left({3\over\hat\varepsilon}+4\right) m +i\d m\cr
+&i{\al\over4\pi}\left\{{p\cdot(\eta+v)\over m^2}
\left[p\cdot (\eta-v)\;\tilde I_2^g+\pn\; \tilde 
I_2^\eta+\vv{}^2\;\pv\; \tilde I_2^v
\right]\right\}\cr
-2&i{\al\over4\pi}\; p\cdot(\eta+v)\left\{
{p\cdot(\eta-v)\over m^2} \left({1\over2} \tilde I_2^p+
\tilde \cI_3^g\right)+
{p\cdot\eta\over m^2} \left({1\over2} \tilde I_2^\eta+\tilde 
\cI_3^\eta\right)+
{\vv{}^2\;p\cdot v \over m^2} \left({1\over2} \tilde 
I_2^v+\tilde \cI_3^v\right)
 \right. \cr
+&
p\cdot(\eta-v)\; \tilde I_3^{pp}+{(\pn)^3\over m^2}\tilde 
I_3^{\eta\eta}+\vv{}^2
{(\pv)^3\over m^2}\tilde I_3^{vv}+
\pn\left[1+{\pn\over m^2}\;p\cdot(\eta-v)\right] \tilde I_3^{p\eta}\cr
+&\left.
\pv\left[\vv{}^2+{\pv\over m^2}\; p\cdot(\eta-v)\right] \tilde 
I_3^{pv}+
{\pn\; \pv\over m^2} (\pv+\vv{}^2\; \pn)\; \tilde I_3^{\eta v}
 \right\}\,,
}
\eqn\theshifter
$$
we refer to the appendix
for the exact meaning of the additional notation here. Recall that only
the mass shift counterterm appears in Eq.~\theshifter.
The first term on the R.H.S.\ here arises from the first term on the
R.H.S.\ of (\selfE)
which is the integrand of the self-energy in Feynman
gauge. The gauge invariance of $\d m$ means that this is
the correct answer. We need to see that the other terms all cancel on
shell no matter what exact on-shell point is employed.
Using the equations (63) and (64) from the appendix, we can
see that they do and that we obtain the standard result:
$$
\d m=m\frac{\a}{4\pi}\( \frac3{\hat \varepsilon} +4\)\,,
\eqn\no
$$
where $\dfrac1{\hat\varepsilon}=
\dfrac1{2-\omega}-\gamma_{\rm E}+{\rm ln}4\pi$.

To verify that the infra-red singularities cancel we should consider
$\bar\Delta$, which we recall is where they arise. We find the
following terms containing infra-red divergences:
$$
\eqalign{
m^2\bar\De_{IR}=i{\al\over4\pi} \int_0^1 du\; u^{2\om-5}\big\{-2&
\big.
+2\int_0^1 \Dx \left[1+\vv{}^2-2\vv{}^2 x\right] \cr
-&
\left.(1-\vv{}^2)\int_0^1 \Dx\; x{3+\vv{}^2-2\vv{}^2 x\over 2
(1-\vv{}^2 x)}
\right\}
}
\,,
\eqn\no
$$
where the subscript \lq IR\rq\ signifies that only the infra-red
singular terms have been retained. The first term comes from the
covariant part of the self-energy and the others have a non-covariant
origin. We find it remarkable, and highly
gratifying, that the sum of the integrals over $x$ gives just $+2$
and so {\it we see that
there is no infra-red divergence in the dressed propagator}.

Since this is the main result of this paper let us stress that we do
not see any {\it a priori} reason
why these divergences should cancel --- other
than our original motivation. It is certainly {\it not} the case that
they cancel for any point on the mass shell. We have verified this
by changing the relative
sign of the vector ${\bold v}$ between the dressing (\thedressing)
and the choice of mass shell point, (\massShellCond). The
infra-red divergences did not then cancel. This shows the great
sensitivity of the calculation.

For completeness we now give the full expressions for $Z_2$ and $Z'$.
We found
$$
Z_2=1+{\al\over4\pi}\left\{{1\over\hat\varepsilon}\left[
3+2\chi({\bold v})\right]-4(1-\vv{}^2)\chi({\bold v}) 
-4\,\kappa({\bold v})\right\}\,,
\eqn\no
$$
and
$$
Z'={\al\over4\pi}\left\{{1\over\hat\varepsilon}\left[
{1\over\vv{}^2}+{1+\vv{}^2\over2\vv{}^2}\chi({\bold v})\right]-
{1\over\vv{}^2}(1-\vv{}^2)\chi({\bold v})-{1+\vv{}^2\over \vv{}^2}\,
\kappa({\bold v})\right\}\,,
\eqn\no
$$
where
$$
\kappa({\bold v})={1\over|\vv|}\Big[L_2(\vert
\vv\vert)-L_2(-\vert\vv\vert)\Big]\,,
\eqn\no
$$
where $L_2$ is the dilogarithm ($L_2(x)
=-\int_0^xdt/t\;{\rm ln}[1-t]$). 
In the small ${\bold v}$ limit these reduce to the expressions we found
in Ref.\ \LETTER, which in turn reduce to the Coulomb gauge
result\up{\ADKINS,\MONSTER} for ${\bold v}\to 0$. We have also checked
that these agreements hold for the results for the
individual functions, $\a,\cdots,\eps$. (Although to compare with the
results of  Ref.\ \LETTER\ for infra-red divergent terms, one needs to
make the translation: $1/\hat\varepsilon\to{\rm ln}\lambda^2/m^2$, where
$\lambda$ is a small photon mass.)
These limits provide a further check upon our results.

\bigskip\bigskip\goodbreak
\ni {\bf 4) Conclusions}
\smallskip
\ni We have seen that the electron propagator is infra-red finite in
the class of gauges (\dgauge) if a suitable on-shell condition is
used. This calculation may also be understood as the
calculation of a dressed propagator in a general gauge.
The renormalisation procedure was completely standard except for the
matrix nature of the fermion wave-function renormalisation. This was
introduced in Ref.\ \LETTER\ and appears rather natural given the
subtleties concerning boosting charged states.
We stress again that the cancellation of the various infra-red
divergences that appear in the individual terms is not fortuitous
but has been predicted in Ref.'s \MONSTER\ and \LETTER. We believe that
this is compelling evidence that the description of an asymptotic
electron which we employ has a firm physical basis. Using 
Ref.~\JACKSOL\  it may be seen that the soft divergences will 
exponentiatiate and so we expect these results to hold at all orders. 
We also stress that we have calculated the wave function 
renormalisation constants explicitly and that they may be used to 
find S-matrix elements involving incoming and outgoing dressed charges. 

Our requirement of the particular renormalisation point used in this
paper makes it clear that gauge invariance alone does not
provide an infra-red finite propagator. We have tried to stress here
the need for an understanding of what meaning (if any!) a
gauge-invariant dressed field possesses. The dressings we have studied
correspond to velocity eigenstates. Other types of dressings should,
we feel, also be constructed and investigated.

As far as the further applications of the dressed fields of this
paper are concerned, the extension of this approach
to the electron-photon vertex
functions is the obvious
next step. If the momentum transfer is non-zero the incoming and
outgoing electrons will have different velocities and should
accordingly be differently dressed, we therefore do not expect the
infra-red divergences present in the usual, undressed
vertex to cancel in any particular
gauge, since no gauge condition would remove all the dressings.
However, if we keep the dressings we expect the dressed vertex to
be infra-red finite in any gauge if the appropriate mass shell 
conditions 
for the fermions are chosen. These calculations will be presented 
elsewhere. 

\bigskip
As far as QCD is concerned, it is clearly harder to construct gauge 
invariant descriptions of charges. In perturbation theory, 
dressings for quarks and gluons have been constructed and 
shown to give a gauge-independent meaning to the
concept of colour charges\up{\COLOUR}. It has also been
seen that there is
an obstruction to dressing colour charges non-perturbatively\up{\PANP}.
A proof of this, a treatment of perturbative dressings for quarks and 
gluons in QCD and a full discussion of the implications of
these matters is to be found in Ref.~\MONSTER. We also refer to 
Ref.'s \QCDONE-\QCDLAST. For theories where the
gauge symmetry is spontaneously broken, dressings may be constructed in
the Higgs sector\up{\SSB}. Perturbative and non-perturbative 
studies of dressed, non-abelian Green's functions have, we feel, 
many practical applications. 

\bigskip\goodbreak
\ni {\bf Acknowledgements:} EB thanks the HE group at BNL for its warm
hospitality while this work was being completed 
and DGICYT for financial support.
MJL thanks project CICYT-AEN95-0815 for
support, E. d'Emilio and L. Lusanna for hospitality and discussions in 
Pisa and Florence and M.\ Stingl for correspondence.

\bigskip\bigskip\goodbreak
\ni {\bf Appendix: About the Integrals}
\smallskip
\ni A treatment of integrals required for calculations 
in Coulomb gauge may be found in Ref.\ 
\ADKINS. The integrals considered here are related to that discussion,  
but are more general in that an extra vector is involved in our case.

\ni \underbar{i) General Formulae}
\smallskip
\ni We need the generic integral
$$
\integ {1\over (k^2-2k\cdot p -M^2)^\al}{k_{\mu_1}\cdots k_{\mu_n}\over
[\noncov]^\be }\,,
\eqn\no
$$
where the second factor in the denominator reflects the structure of 
the gauge boson propagator, (\dgaugeprop). 
We first go to Euclidean space and exponentiate the denominators using
$$
X^{-\al}=
{1\over \Gamma(\al)}  \int_0^\infty dy\; y^{\al-1} {\rm e}^{-X y}\,.
\eqn\no
$$
Then we make use of 
$$
\int_{\rm Eucl}      {   d^{2\om} k\over  (2\pi)^{2\om}    } 
{\rm e}^{-[k^t{\cal M} k - J^t k]}    
={\pi^\om\over(2\pi)^{2\om}}{1\over\sqrt{\det {\cal M}}}\; {\rm e}^{{1\over4}
J^t {\cal M}^{-1} J}\,.
\eqn\no
$$
In our case the $(2\om)\times(2\om)$ matrix ${\cal M}$ is
$$
{\cal M}_{\mu\nu}=(y+z) 
\de_{\mu\nu}-z \eta_\mu\eta_\nu - z v_\mu v_\nu\,,
\eqn\no 
$$
where $y$ and $z$ are the Feynman parameters 
used to exponentiate the two 
denominators in our generic integral. Similarly in our case 
$$
J_\mu=2 y\; p_\mu\,.
\eqn\no
$$
To go from scalar integrals to vector or tensor ones, we simply have to 
take derivatives according to the recipe
$$
k_\mu \rightarrow {1\over 2y} {\p\over\p p^\mu}\,.
\eqn\no
$$
Upon changing the variables
$$
y=(1-x) t;\quad z= x t;\qquad \Rightarrow\quad
 dy\,dz=t\;dx\,dt;\qquad x\in[0,1];\quad
t\in[0,\infty]\,.
\eqn\no
$$
we get
$$
\det {\cal M}= t^{2\om} (1-x)(1-\vv^2 x)\,,
\eqn\no
$$
and so, back in Minkowski space, we have 
$$
A_{\mu\nu}\equiv [{\cal M}^{-1}]_{\mu\nu}=g_{\mu\nu}+{x\over 1-x}
\eta_\mu\eta_\nu - {x\over 1-\vv{}^2 x} v_\mu v_\nu\,.
\eqn\Izero
$$
One finally thus obtains
$$
\eqalign{
\integ & {1\over (k^2-2k\cdot p -M^2)^\al}{B\over
[\noncov]^\be }\cr
&\qquad\quad
={(-1)^{\a+\b} 
\over(2\pi)^{2\om}}{i \pi^\om \over \Gamma(\al)\Gamma(\be)}
\int_0^1 {dx\over \sqrt{1-x}\sqrt{1-\vv{}^2 x}} (1-x)^{\al-1} 
x^{\be-1} C\,,
}
\eqn\BCrel
$$
where various pairs of $B$'s and $C$'s are related as follows:
$$
\eqalign{
B= &1\,,\qquad\,\, C= \frac{\Gamma(\a+\b-\omega)}{\Delta_g^{\a+\b-\omega}} 
\,,\cr
B= &k_\mu\,,\quad\,\,\,\,\, C= (1-x) 
\frac{\Gamma(\a+\b-\omega)}{\Delta_g^{\a+\b-\omega}} \(Ap\)_\mu 
\,,\cr
B= &k_\mu k_\nu\,,\,\quad C= (1-x)^2\(Ap\)_\mu \(Ap\)_\nu 
\frac{\Gamma(\a+\b-\omega)}{\Delta_g^{\a+\b-\omega}}
-\frac12 A_{\mu\nu} 
\frac{\Gamma(\a+\b-1-\omega)}{\Delta_g^{\a+\b-1-\omega}}
\,,}
\eqn\no
$$
and lastly
$$
\eqalign{
B=k_\mu k_\nu k_\rho\,,\quad C=& (1-x)^3 \(Ap\)_\mu
 \(Ap\)_\nu  \(Ap\)_\rho 
\frac{\Gamma(\a+\b-\omega)}{\Delta_g^{\a+\b-\omega}}\cr 
&-\frac{(1-x)}2\left[ A_{\mu\nu}  \(Ap\)_\rho + 
 A_{\mu\rho}  \(Ap\)_\nu +
 A_{\nu\rho}  \(Ap\)_\mu \right]
\frac{\Gamma(\a+\b-1-\omega)}{\Delta_g^{\a+\b-1-\omega}} 
\,,}
\eqn\no
$$
where we have further introduced the notation
$$
\Delta_g=(1-x)\left[ (1-x)p_\mu p_\nu A^{\mu\nu}+M^2\right]
\,.
\eqn\no
$$

We also use the relation 
$$
{1\over k^2}{1\over\cov}=
\int_0^1 {du\over [k^2-2 u k\cdot p -u (m^2-p^2)]^2}\,,
\eqn\no
$$
to, where necessary, combine the two covariant denominators coming from 
the fermion propagator and the vector boson propagator.

For integrals with one 
or two covariant denominator structures $\Delta_g$ takes on different 
forms. For an integral with one covariant and one 
non-covariant denominator term (so two structures in total) 
we have for $\Delta_g$
$$
\De_2=(1-x)(\Pi+m^2-p^2)\,,
\qquad \Pi=(1-x)p^2+x (p\cdot\eta)^2-{(1-x)x\over
1-\vv{}^2x} (p\cdot v)^2\,. 
\eqn\Dtwo
$$
If we have 
two non-covariant structures and one non-covariant term in the 
denominator, then we have for $\Delta_g$
$$
\Delta_3=u(1-x)\left\{ u\Pi+m^2-p^2\right\}\,,
\eqn\no
$$
the similarity between these last two equations indicates the utility 
of this notation. 

\bigskip\goodbreak
\ni \underbar{ii) The On-Shell Integrals Needed for the Mass Shift}
\smallskip\ni To compute the mass shift,
we need to know the following integrals for $p^2=m^2$ and arbitrary 
$p\cdot\eta$, $p\cdot v$, $v$:
$$
{16\pi^2\over {i(m^2)^{\omega-2}}}
\integ {1\over\cov} {k_\mu \over\noncov} =
I_2^p p_\mu+ p\cdot\eta\; I_2^\eta \eta_\mu+ p\cdot v \;  
I_2^v v_\mu \,,
\eqn\no
$$
where we define for on-shell momentum, $p$
$$
\eqalign{
\tilde 
I_2^p&=\int_0^1 {dx\over \sqrt{1-x}\sqrt{1-\vv{}^2 x}} (1-x)
   \left[{1\over\hat\varepsilon}-\log \frac{\tilde\De_2}{m^2} \right] \,,\cr
\tilde I_2^\eta&=\int_0^1 {dx\over \sqrt{1-x}\sqrt{1-\vv{}^2 x}} x
   \left[{1\over\hat\varepsilon} 
 -\log \frac{\tilde\De_2}{m^2} \right]\,, \cr
\tilde I_2^v&=\int_0^1 {dx\over \sqrt{1-x}\sqrt{1-\vv{}^2 x}}
 {-x(1-x)\over 1-\vv{}^2x}
   \left[{1\over\hat\varepsilon} 
-\log \frac{\tilde\De_2}{m^2} \right]\,, \cr
}
\eqn\no
$$
and, as in the main body of the paper, a tilde signifies that the 
function is evaluated on an arbitrary point on the 
mass shell, $p^2=m^2$. 

We also need the integrals 
$$
\eqalign{
{16\pi^2\over{i{(m^2)^{\omega-2}}}} 
& \integ \frac1{k^2} {1\over\cov}{k_\mu k_\nu \over\noncov}=
I_3^g\; g_{\mu\nu} + I_3^\eta\; \eta_\mu\eta_\nu \cr& 
+ I_3^v\; v_\mu 
v_\nu 
 +I_3^{pp}\; 
p_\mu p_\nu + (p\cdot\eta)^2\, I_3^{\eta\eta}\; \eta_\mu\eta_\nu
 + (p\cdot v)^2 \, I_3^{vv}\; v_\mu v_\nu 
 \cr & +
p\cdot\eta\; I_3^{p\eta} (p_\mu\eta_\nu+\eta_\mu p_\nu) +
p\cdot v\; I_3^{pv} (p_\mu v_\nu+v_\mu p_\nu)
+p\cdot\eta\; p\cdot v\;  I_3^{\eta v} (v_\mu\eta_\nu+\eta_\mu v_\nu),
}\eqn\no
$$
where
$$
\eqalign{
I_3^g&={1\over2} I_2^p-
\int_0^1 du
\;\log u  \int_0^1 {dx\over \sqrt{1-x}\sqrt{1-\vv{}^2 x}}(1-x)
={1\over2} I_2^p+\cI_3^g
\,, \cr
I_3^\eta&={1\over2} I_2^\eta-
\int_0^1 du\;\log u \int_0^1 {dx\over \sqrt{1-x}\sqrt{1-\vv{}^2 x}} x
={1\over2} I_2^\eta+\cI_3^\eta \,, \cr
I_3^v&={1\over2} I_2^v-
\int_0^1 du\;\log u \int_0^1 {dx\over \sqrt{1-x}
\sqrt{1-\vv{}^2 x}}{-x (1-x)
\over 1-\vv{}^2 x}={1\over2} I_2^v+\cI_3^v \,, }
\eqn\no
$$
and we see that the $u$ integral is just $-1$; 
similarly for on-shell $p$ we have 
$$
\eqalign{
\tilde I_3^{pp}&=
\int_0^1 du \int_0^1 {dx\over \sqrt{1-x}\sqrt{1-\vv{}^2 x}}
{-(1-x)^2\over\tilde\Pi} \,, \cr
\tilde I_3^{\eta\eta}&=
\int_0^1 du \int_0^1 {dx\over \sqrt{1-x}\sqrt{1-\vv{}^2 x}}
{-x^2\over\tilde\Pi}\,, \cr
\tilde I_3^{vv}&=
\int_0^1 du \int_0^1 {dx\over \sqrt{1-x}\sqrt{1-\vv{}^2 x}}
{-x^2 (1-x)^2\over(1-\vv{}^2 x)^2}{1\over\tilde\Pi} \,, \cr
\tilde I_3^{p\eta}&=
\int_0^1 du \int_0^1 {dx\over \sqrt{1-x}\sqrt{1-\vv{}^2 x}}
{-x (1-x)\over\tilde\Pi} \,, \cr
\tilde I_3^{pv}&=
\int_0^1 du \int_0^1 {dx\over \sqrt{1-x}\sqrt{1-\vv{}^2 x}}
{x (1-x)^2\over1-\vv{}^2x}{1\over\tilde\Pi} \,, \cr
\tilde I_3^{\eta v}&=
\int_0^1 du \int_0^1 {dx\over \sqrt{1-x}\sqrt{1-\vv{}^2 x}}
{x^2 (1-x)\over1-\vv{}^2x}{1\over\tilde\Pi} \,.
}\eqn\no
$$
and the trivial $u$ integral just yields $1$. 
It takes some algebra to show that
$$
\eqalign{
&p\cdot(\eta-v)\; \tilde I_3^{pp}+
{(\pn)^3\over m^2}\tilde I_3^{\eta\eta}+\vv{}^2
{(\pv)^3\over m^2}\tilde I_3^{vv}+
\pn\left[1+{\pn\over m^2}\;p\cdot (\eta-v)\right]\tilde I_3^{p\eta}\cr
&+\pv\left[\vv{}^2+{\pv\over m^2}\;p\cdot (\eta-v)\right]\tilde 
I_3^{pv}+
{\pn\; \pv\over m^2} (\pv+\vv{}^2 \pn)\; \tilde I_3^{\eta v} \cr
&=-{1\over m^2} \int_0^1 {dx\over \sqrt{1-x}\sqrt{1-\vv{}^2 x}}
\left\{p\cdot\eta -p\cdot v
{1-x\over 1-\vv{}^2 x}\right\}\,, 
}
\eqn\Ione 
$$
and similarly that 
$$
{p\cdot(\eta-v)\over m^2}\tilde \cI_3^g
+{p\cdot\eta\over m^2}\;\tilde 
\cI_3^\eta+\vv{}^2 {p\cdot v\over m^2}\;\tilde \cI_3^v
={1\over m^2} \int_0^1 {dx\over \sqrt{1-x}\sqrt{1-\vv{}^2 x}}
\left\{p\cdot\eta -p\cdot v
{1-x\over 1-\vv{}^2 x}\right\}\,,
\eqn\Itwo 
$$
but armed 
with these results we may easily obtain the standard result for the 
mass shift, as given by (\theshifter).  
\bigskip
\goodbreak
\ni \underbar{iii) An Example}
\smallskip\ni 
We now round off this appendix by showing how the above general 
discussion may be applied to compute a particular non-covariant 
integral. Consider therefore 
$$
\eqalign{
\frac1{(m^2)^{\omega-2}}\integ\frac1\cov & \frac1\noncov=\cr 
&\qquad 
\frac{i\pi^\omega}{(2\pi)^{2\omega}}\int^1_0 \!
\frac{dx}{\sqrt{1-x}\sqrt{1-{\bold v}^2 x} } 
\frac{\Gamma(2-\omega)}{\({\displaystyle
{{\Delta_2}/{m^2}}}\)^{2-\omega}}\,,
}\eqn\no
$$
where $\Delta_2$ is given in (\Dtwo). This relation 
follows from (\BCrel). We now expand this in $\varepsilon
=(2-\omega)$ and obtain  
$$
\frac i{16\pi^2}
\int^1_0 \!
\frac{dx}{\sqrt{1-x}\sqrt{1-{\bold v}^2 x} } 
\Big\{
\frac1{\hat\varepsilon}-{\rm ln}\Delta_2
\Big\} +O(\varepsilon)\,.
\eqn\eXpnder
$$
The change of variables, $x=(1-t^2)/(1-{\bold v}^2t^2)$, is now 
useful. The integral coefficient of the pole in $\varepsilon$ 
can then be re-expressed as
$$
\eqalign{
\int^1_0 \!
\frac{dx}{\sqrt{1-x}\sqrt{1-{\bold v}^2 x} } =& 
2\int^1_0\!dt\frac1{1-{\bold v}^2t^2}\cr
=&\frac1{\vert {\bold v}\vert} {\rm ln} \frac{1-\vert 
{\bold v}\vert}{1+\vert {\bold v}\vert }\equiv-\chi({\bold v})\,,
}
\eqn\no
$$
where we recall the definition of $\chi$ from (\chiEq).

The second integral in (\eXpnder) depends on $p$. We will not calculate 
it for an arbitrary $p$, but rather in a Taylor expansion around the 
correct, physical pole for the dressing we use. 
Again employing the notation that bars over 
functions signify that they are evaluated at $p=m\gamma(\eta+v)$, we 
find
$$
\bar\Pi=\frac{m^2}{1-{\bold v}^2x}\,,\quad
\bar\Delta_2 =(1-x) \frac{m^2}{1-{\bold v}^2x}\,,\quad\left. 
\frac{\pa}{\pa p^2} 
\Delta_2\right\vert_{p=m\gamma(\eta+v)}=-x(1-x)\,.
\eqn\no
$$
Thus we obtain 
$$
\eqalign{
\int^1_0\!dx \frac{{\rm 
ln}({\Delta_2}/{m^2})}{\sqrt{1-x}\sqrt{1-{\bold v}^2}x} = & 
\int^1_0\!dx \frac{{\rm ln}(1-x)-{\rm ln}(1-{\bold v}^2 
x)}{\sqrt{1-x}\sqrt{1-{\bold v}^2}x} \cr &\quad 
-\frac{p^2-m^2}{m^2}
\int^1_0\!dx \frac{x(1-{\bold v}^2x)}{\sqrt{1-x}\sqrt{1-{\bold 
v}^2}x}\,. 
}
\eqn\no
$$
Repeating the transformation of variables, these two integrals yield 
respectively 
$$
\eqalign{
\int^1_0\!dx \frac{{\rm ln}(1-x)-{\rm ln}(1-{\bold v}^2 
x)}{\sqrt{1-x}\sqrt{1-{\bold v}^2}x} =& 
2\int^1_0\!dt \frac{{\rm ln}t^2}{1-{\bold v}^2t^2}\cr
=& \frac{2}{\vert {\bold v}\vert}
\Big[
L_2(-\vert{\bold v}\vert) - L_2(\vert{\bold v}\vert)
\Big] \equiv  -2\kappa({\bold v})
\,,
}
\eqn\no
$$ 
and 
$$
\int^1_0\!dx \frac{x(1-{\bold v}^2x)}{\sqrt{1-x}\sqrt{1-{\bold v}^2}x}
=\frac34-\frac1{4{\bold v}^2}-\frac{(1-{\bold v}^2)(1+3{\bold v}^2)}{8 
{\bold v}^2}\chi({\bold v}) 
\,.
\eqn\no
$$

Putting everything together we obtain for our exemplary integral
$$
\eqalign{
\frac1{(m^2)^{\omega-2}} & \integ  \frac1\cov\frac1\noncov= 
\frac i{16\pi^2}\Bigg\{\Bigg. -\chi({\bold v})
\frac1{\hat\varepsilon} +2\kappa({\bold v})\cr & 
\!\!\!\! +\frac{p^2-m^2}{m^2}\left[ 
\frac34-\frac1{4{\bold v}^2}-\frac{(1-{\bold v}^2)(1+3{\bold v}^2)}{8 
{\bold v}^2}\chi({\bold v}) 
\right] \Bigg.\Bigg\} +O\((p^2-m^2)^2\)
\,.}
\eqn\no
$$
In the limit ${\bold v}\to0$ this correctly yields  
$$
\frac i{16\pi^2}
\left\{\frac2{\hat\varepsilon}+4+\frac{p^2-m^2}{m^2}\frac43\right\}
\,.
\eqn\no
$$
Very similar manipulations yield the other integrals we require. 

Finally 
we should also mention that various consistency relations between 
integrals have been checked (e.g., replacing a factor of 
$(k\cdot\eta)^2$ in a  
numerator by $k^2+{\bold k}^2$ and performing the two resulting 
integrals separately) and seen to hold.

%\vfill\eject
\bigskip\bigskip\goodbreak

  \immediate\closeout\reffile%\parindent=20pt
  \centerline{{\bf References}}\bigskip\eightpoint\frenchspacing%
  \input refs.tmp\vfill\eject\nonfrenchspacing

\bye